# Non-equilibrium Molecular Dynamics Study of Surface Wettability Effects on Pool Boiling of Water over Nanoscale Aluminum Substrate

Farhad Sotoudeh, Jafar Ghazanfarian[*]

Mechanical Engineering Department, Faculty of Engineering, University of Zanjan, University Blv., Zanjan, Iran
[*]Corresponding author's email: j.ghazanfarian@znu.ac.ir

**Abstract**

Non-equilibrium molecular dynamics (NEMD) simulations were used to study pool boiling of water films on an ultra-thin planar aluminum substrate as well as the effect of surface wettability. The simulation geometry is a 10 nm-thick water film on an FCC aluminum substrate heated from 300 K to 900 K. The first peak acceleration onset time of the film, as the measure of the nucleation start, has been observed. The average heating rates of the near-wall water were 0.064, 0.048, and 0.035 K·ps$^{-1}$ for hydrophilic, neutral, and hydrophobic surfaces, respectively. Boiling curves shows that the critical heat flux (CHF) equals 5216, 3979, and 2525 MW·m$^{-2}$ at wall temperatures of 466, 502, and 561 K, respectively. The minimum heat flux (MHF, Leidenfrost point) is equal to 2157, 2463, and 2366 MW·m$^{-2}$ at wall temperatures of 767, 784, and 746 K, respectively. Interfacial HTC remains higher for longer times under the hydrophilic condition, whereas Kapitza resistance is low initially but then increases sharply after transition to film boiling with the highest values for the hydrophobic surface. In general, the results demonstrate that engineering aluminum wettability towards intense hydrophilicity diminishes the explosive boiling point, increases CHF, and enhances nanoscale thermal management performance.

# 1 Introduction

With the rapid advancement of micro/nanotechnologies, thermal management of high-power-density electronic and energy systems is one of the most critical challenges of modern electronics. Boiling plays a crucial role in enhancing the performance and reliability of these systems, particularly at small sizes. Among different modes of boiling, pool boiling is especially effective due to the huge heat transfer rate provided by the latent heat of evaporation [1-3].

Wettability of the solid surface, composition, and surface structure directly influence bubble nucleation and heat transfer rates. Contact angle, a measure of wettability, determines the liquid-solid interaction and can significantly affect boiling behavior. Wettability has a very powerful effect on bubble creation, growth, and detachment and hence has a significant impact on global boiling performance [4,5].

Much work over the last few decades has been conducted on optimizing boiling heat transfer by varying surface properties. Numerous investigations have been able to demonstrate that mixed or biphilic surfaces with wettability formed of hydrophilic and hydrophobic domains enhance nucleation and stabilize bubbles [6-10]. Optimally ordered hydrophilic/hydrophobic domains have been found to suppress stable vapor films and increase the critical heat flux (CHF) [6,9]. Furthermore, geometrically-ordered patterns such as columns, stripes, and grids are proposed to improve further the thermal performance [7,10].

At the nanoscale, the ratio of surface area to volume increases immensely, promoting interfacial phenomena in boiling heat transfer [11]. However, experimental measurement and observation of boiling behavior at this scale are extremely difficult [12]. Here, simulation through molecular dynamics (MD) has been a really strong tool, enabling study of atomistic processes and fundamental knowledge unavailable by experiments [13].

MD-based studies have promoted the understanding of boiling interfacial phenomena. For instance, nanoparticles with tailored wettability can modulate substrate surface energy and enhance boiling performance [14,15]. Graphene coatings have been found to influence bubble nucleation in other studies [16], nanoscale roughness influences bubble growth mechanism [17], and contact angle changes influence bubble generation frequency [18]. In addition, atomistic simulations have indicated that temperature gradients [19], film thickness of liquid [20], and external fields [21] alter boiling behavior significantly.

Earlier, efforts have been made to integrate nanoscale surface roughness with wettability control. Computational simulations indicate that nanogrooves, nanopillars, and hemi-spherical cavities, together with wisely controlled interfacial energy, can highly promote evaporation and suppress the onset of nucleate boiling [22-25]. Highly hydrophilic structures are particularly efficient in offering favorable locations for energy concentrating and bubble nucleation [23,26]. But the literature review suggests that large portions of the earlier experimental and computational investigations have been using working fluids such as argon or substrates such as copper, platinum, and gallium [27-30].

On the contrary, aluminum possesses high thermal conductivity, widespread availability at low cost, and hence is extremely prevalent in industrial and electronic applications. Despite these advantages, pool boiling of water on aluminum surfaces, especially under conditions of varying wettability, has not been studied yet. In this work, non-equilibrium molecular dynamics (NEMD) simulations are employed to investigate pool boiling of water films on smooth aluminum surfaces with three wettability conditions: strongly hydrophilic, neutral, and strongly hydrophobic. In the first stage, the effect of interfacial energy ratio on water contact angle is quantified. Subsequently, the role of wettability in boiling dynamics and its impact on heat flux performance is systematically analyzed.

## 2 Details of Simulations

The boiling process of ultra-thin water films on aluminum surfaces with different wettability conditions has been investigated using classical molecular dynamics (MD) simulations implemented in the *Large-scale Atomic/Molecular Massively Parallel Simulator* (LAMMPS) [31]. The resulting trajectory data and atomic configurations were post-processed and visualized using the *Open Visualization Tool* (OVITO) [32].

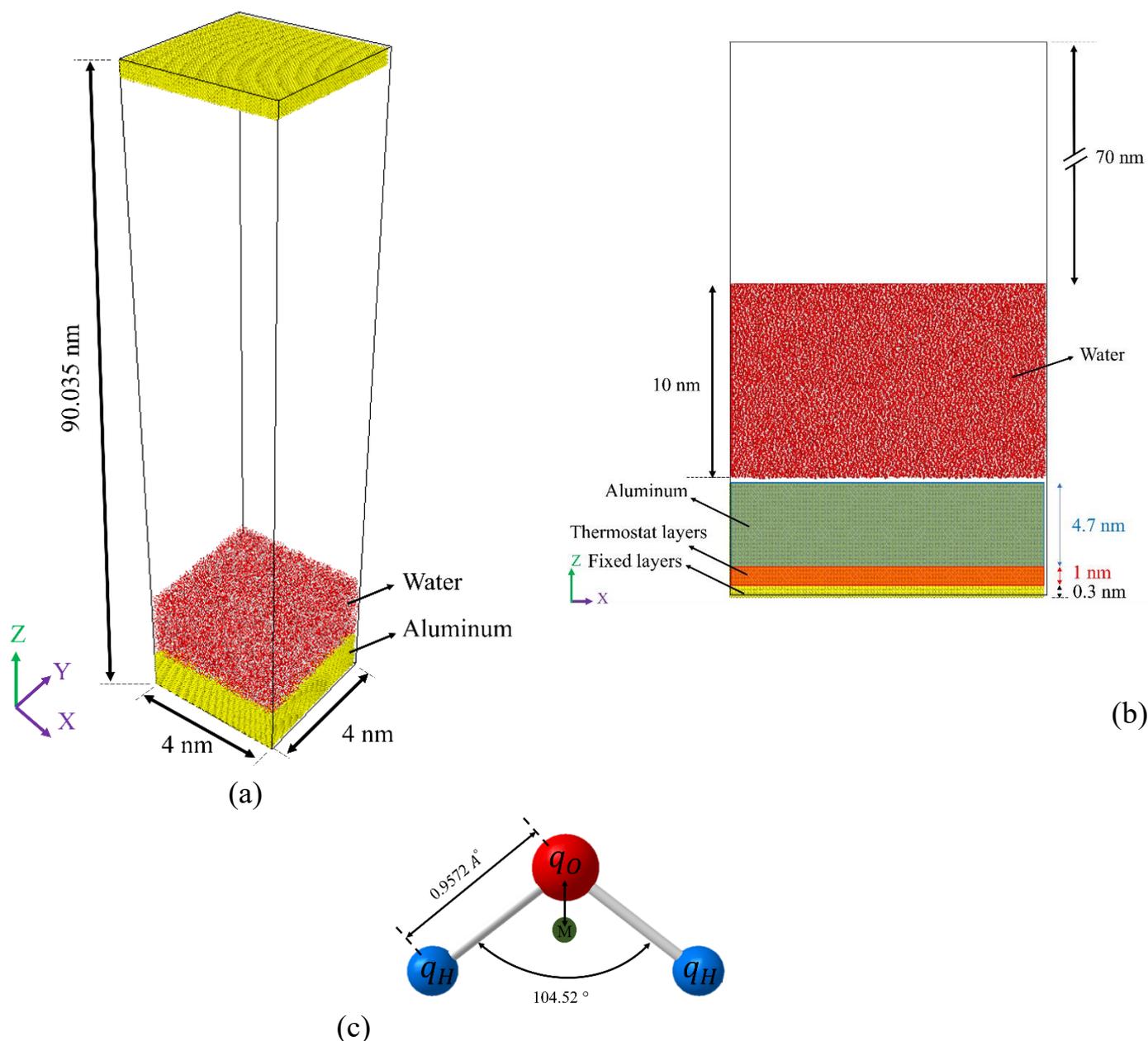

**Figure 1.** (a) Three-dimensional schematic view of the simulated system. (b) Side view illustrating the arrangement of the aluminum substrate and the thin water film. (c) Molecular structure of water represented using the TIP4P model.

A schematic representation of the simulation system is shown in Fig. 1. The dimensions of the simulation box were maintained the same at 4×4×90 nm³ (x×y×z) for all cases. The solid support consisted of aluminum atoms in a face-centered cubic (FCC) crystal structure with lattice constant L = 4.05 Å, calculated from the experimental density of aluminum (2.7 g·cm⁻³). The total thickness of the solid layer in the simulation model was varied to 6 nm along the vertical (z) axis [3,5,21].

To maintain a physically realistic thermostat, the solid substrate was divided into three distinct regions:

1. Bottom layer (thickness 0.3 nm): constrained in order to prevent atomic drift out of the simulation box.

2. Thermostat layer (thickness 1 nm): employed as a heat source, where thermal control was applied.

3. Solid atoms: served as real substrate atoms that transferred heat from the source region to the liquid phase.

The working fluid is water, placed in proximity to the surface of the aluminum substrate. This position corresponds to a density of 1 g·cm$^{-3}$ at 300 K and 1 bar. The film thickness is 10 nm and formed by a total of 124,317 water molecules in all simulations. Periodic boundary conditions have been implemented in the x and y directions, and fixed boundaries in the z direction. The surface of the simulation domain is a reflecting wall.

For describing molecular interactions, the TIP4P water model is employed (see Fig. 1c), which exhibits better agreement with experimental thermophysical properties [33]. The partial charges of hydrogen and oxygen atoms were set to +0.520e and -1.040e, respectively, with an O-H bond length of 0.9572 Å and an H-O-H angle of 104.52°. The Lennard-Jones (L-J) 12-6 potential was adopted to describe Al-Al and Al-O interactions [18]:

$$U_{ij} = 4\varepsilon_{ij}\left[\left(\frac{\sigma_{ij}}{r_{ij}}\right)^{12} - \left(\frac{\sigma_{ij}}{r_{ij}}\right)^{6}\right] \tag{1}$$

where $r_{ij}$ denotes the interatomic distance between particles i and j in Å, while σ and ε represent the zero-potential separation distance (Å) and the potential well depth (eV), respectively. The Lennard-Jones potential is truncated using a spherical cut-off radius of $r_c = 10$ Å. For the Al-O interactions, the Lorentz-Berthelot combining rule is employed to evaluate the cross-interaction parameters $\varepsilon_{Al-O}$ and $\sigma_{Al-O}$ as follows [34]:

$$\sigma_{Al-O} = \frac{\sigma_{Al}+\sigma_O}{2}, \text{ and } \varepsilon_{Al-O} = \alpha\sqrt{\varepsilon_{Al}\varepsilon_O} \tag{2}$$

where α serves as an energy scaling factor to tune the strength of hydrophilic and hydrophobic interactions. For water-water interactions, a long-range pairwise Coulombic term must be incorporated into Eq. (1) [18]:

$$U_{ij} = \frac{C q_{e,i} q_{e,j}}{r_{ij}} + 4\varepsilon_{ij}\left[\left(\frac{\sigma_{ij}}{r_{ij}}\right)^{12} - \left(\frac{\sigma_{ij}}{r_{ij}}\right)^{6}\right] \qquad (3)$$

where C is the Coulomb constant, and $q_{e,i}$ and $q_{e,j}$ represent the charges of the interacting particles. The parameters associated with the interatomic interactions employed in the simulations are summarized in Table 1.

Table 1. parameters associated with the interatomic interactions employed in the simulations

| Particles i, j | α | $q_e$ (e) | $\varepsilon_{ij}$ (kcal/mol) | $\sigma_{ij}$ (Å) |
|---|---|---|---|---|
| H-H | - | +0.520 | 0 | 0 |
| O-O | - | -1.040 | 0.16138 | 3.1650 |
| AL-AL | - | - | 9.04376 | 2.620 |
| AL-H | - | - | 0 | 1.310 |
| AL-O | 1 | - | 1.20809 | 2.8925 |

To investigate the contact angle and surface wettability, a system of a spherical droplet with diameter of the order 10 nm in contact with a solid substrate is simulated. The size of the substrate, as shown in Fig. 2, is set at 25×25×2 nm in the x, y, and z directions, respectively. For ease of constructing the first droplet structure, a cubic water cluster with a side of 8.1 nm has been initially generated, which eventually transforms into a spherical droplet. The simulation is carried out in two successive steps. For the first step, the system is simulated in the NVT ensemble at 300 K for 0.5 ns to reach a state of equilibrium. At the second stage, the statistical ensemble is switched to NVE, and the simulation is again continued for 1 ns to study the dynamic behavior of the droplet. The equilibrium contact angle of the droplet on the solid surface is calculated at the end of this stage and is employed as a quantitative descriptor for the description of surface wettability.

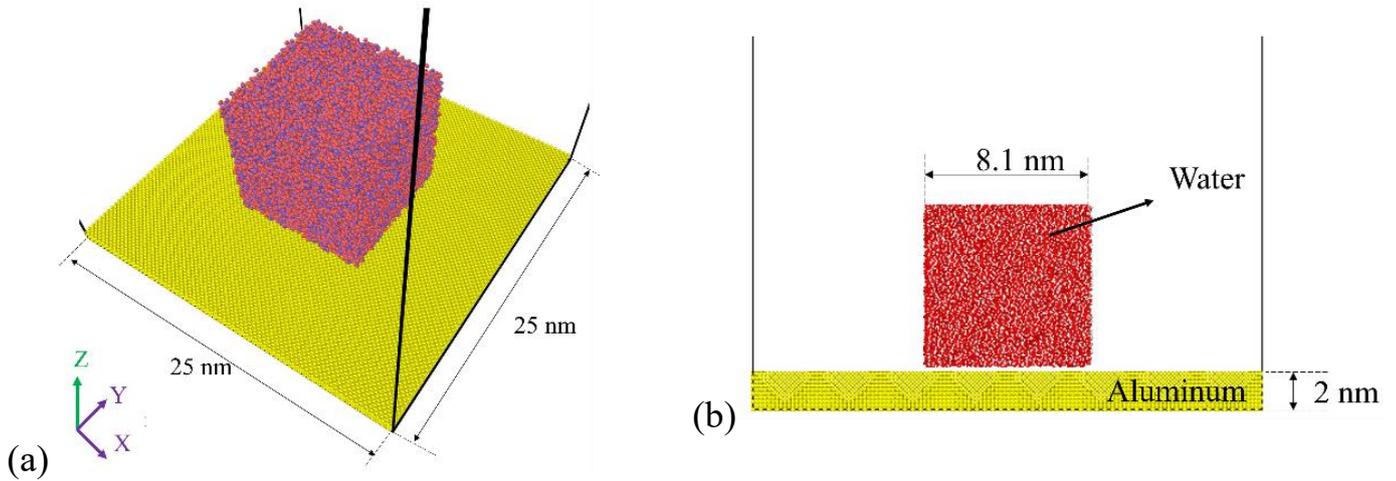

**Figure 2.** Initial configuration of a cubic water droplet prior to relaxation: (a) three-dimensional view, (b) side view.

## 3 Validation and Simulation Procedure

The simulation has been carried out in two stages. In the equilibrium molecular dynamics (EMD) stage, the fluid is equilibrated using the Berendsen thermostat, while the solid substrate is controlled with the Langevin thermostat at a temperature of 300 K. The equilibration process for each component is performed for 3 ns with a time-step of 1 fs, during which the system's temperature and total energy are continuously monitored and recorded [35]. Once these parameters reached a steady state, as illustrated in Figure 3, the system is considered to be in equilibrium [6, 8, 15, 17, 21, 24, 36-39].

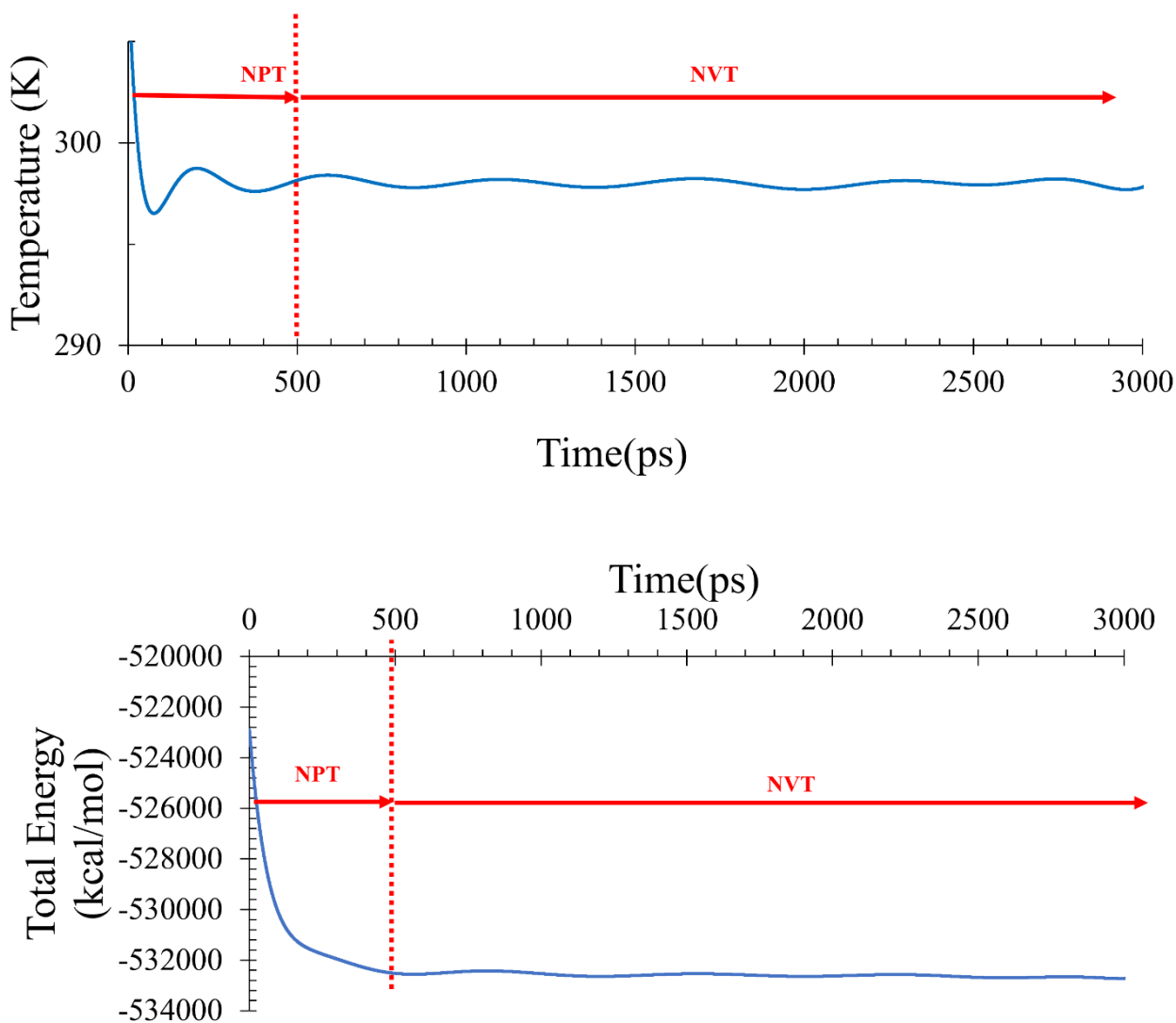

**Figure 3.** Evolution of temperature (the first row) and total energy (the second row) during the equilibration stage, confirming that the system has reached a stable equilibrium state.

In the non-equilibrium molecular dynamics (NEMD) modeling, the thermostat is applied to the intermediate solid substrate layers only, whereas the remaining atoms are simulated in the NVE ensemble. The temperature of the heat source is increased linearly from 300 K to 900 K over a period of 6 ns to mimic the boiling conditions, at a heating rate of 0.1 K·ps$^{-1}$.

After equilibrium has been reached in the system, the radial distribution function (RDF) of water molecules is used to validate the computational strategy adopted for the calculation of the target structure. Figure 4 represents the RDF of oxygen atoms of water molecules. The agreement with literature results was good when comparing results with previous data [40]. The RDF of an atomic sample is the spatial average disposition of atoms relative to each other under the given initial conditions. The similarity with the observed results implies structural disposition of water molecules and presents a test of validity of the approach used in the present study. This similarity with literature guarantees that spatial distributions and molecular interactions have been accurately reflected in the selected force field and simulation conditions. Such validation provides the foundation for the reliability of all the following thermophysical analyses that were performed in this research.

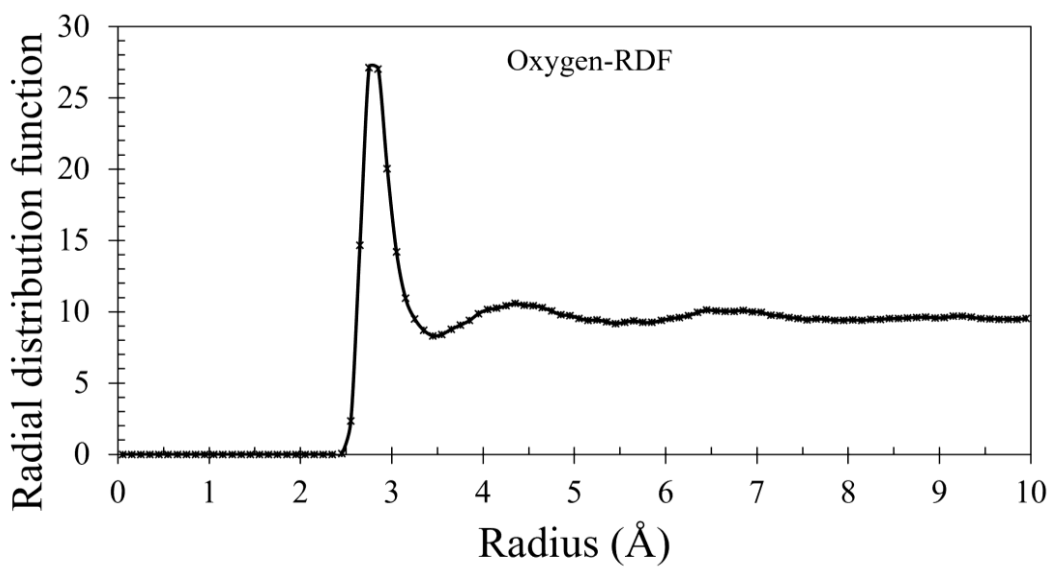

**Figure 4.** Radial distribution function (RDF) of oxygen atoms in water molecules after equilibration of the structure at $T = 300$ K (after 3 ns).

## 4 Results and Discussion

### 4-1 Effect of Surface Potential Energy on Wettability

The contact angle on an ideal flat surface is one of the key parameters for characterizing the wettability between a droplet and a substrate. In this study, to determine the contact angle, a water droplet with a diameter of approximately 10 nm has been placed on an aluminum substrate. Different values of the surface-water interaction parameter (α) were obtained by adjusting the Lennard-Jones potential parameters between surface atoms and water molecules. After performing equilibration under the NVT ensemble and subsequent molecular dynamics simulations under the NVE ensemble, the equilibrium droplet profile was extracted. The contact angle was then determined by fitting a circular curve to the droplet profile and calculating the angle between the tangent line at the contact point and the solid surface.

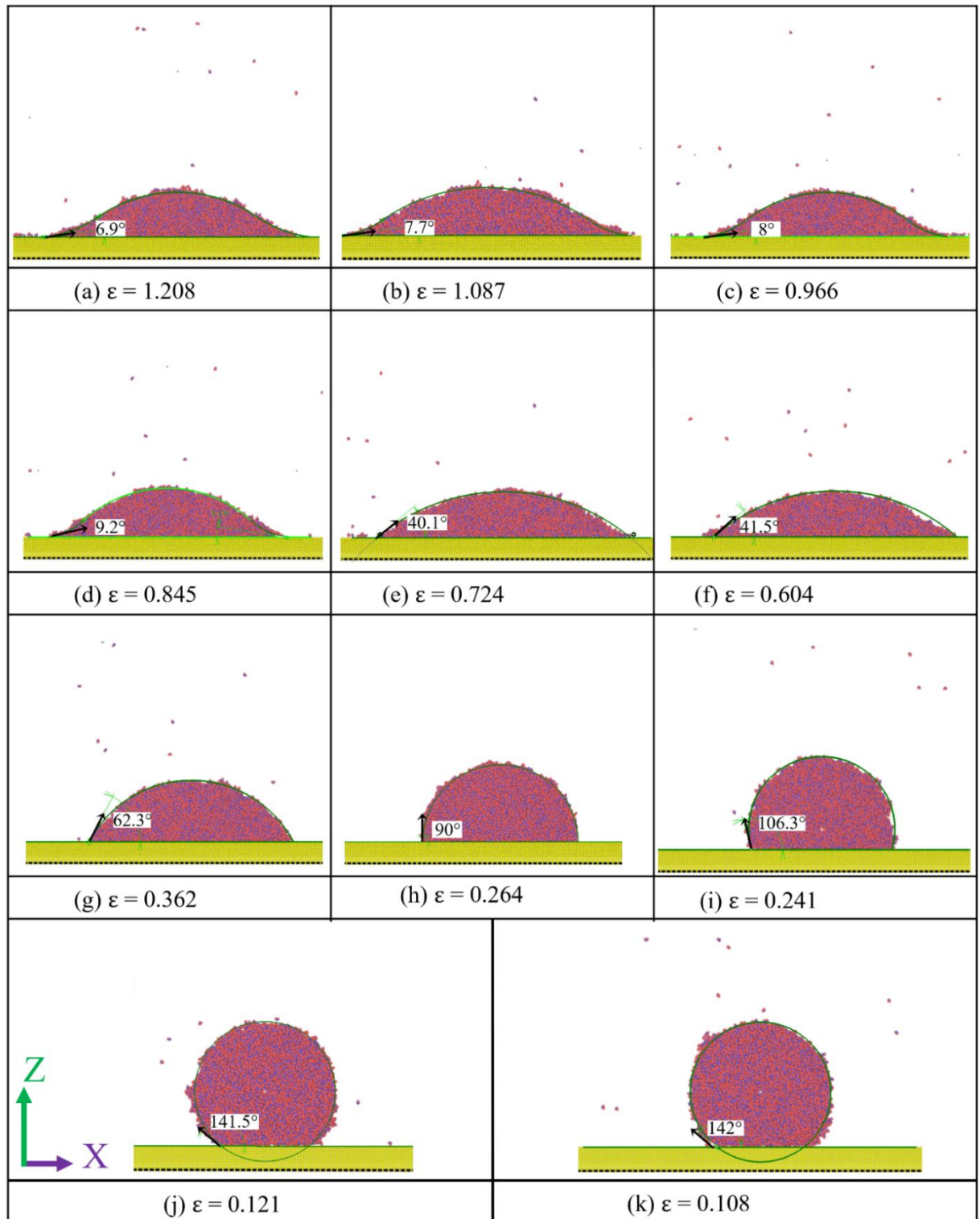

**Figure 5.** Equilibrium configurations of water droplets on surfaces with different wettability conditions.

The findings for substrates with different wettability are presented in Figure 5, demonstrating a clear transition of wetting behavior as the parameter α decreases from fully hydrophilic to very hydrophobic conditions. The precise values of α and the respective equilibrium contact angles are tabulated in Table 2. As depicted, reducing α from around 1 to 0.09 causes the contact angle to change from a highly hydrophilic state (≈6.9°) to an extremely hydrophobic regime (≈142°). Additionally, a neutral wetting surface is realized at α ≈ 0.219, with the contact angle being 90°.

Table 2. Precise values of α and respective equilibrium contact angles.

| Wettability | α | $\varepsilon_{AL-O}$ ($\frac{kcal}{mol}$) | Contact angle° |
|---|---|---|---|
| Strong hydrophilicity | 1 | 1.208 | 6.9 |
| Strong hydrophilicity | 0.9 | 1.087 | 7.7 |
| Strong hydrophilicity | 0.8 | 0.966 | 8 |
| Strong hydrophilicity | 0.7 | 0.845 | 9.2 |
| Hydrophilicity | 0.6 | 0.724 | 40.1 |
| Hydrophilicity | 0.5 | 0.604 | 41.5 |
| Weak hydrophilicity | 0.3 | 0.362 | 62.3 |
| Neutral | 0.219 | 0.264 | 90 |
| Hydrophobicity | 0.2 | 0.241 | 106.3 |
| Hydrophobicity | 0.1 | 0.121 | 141.5 |
| Hydrophobicity | 0.09 | 0.108 | 142 |

This contact angle versus α trend is illustrated in Figure 6. The graph shows an unmistakable nonlinear inverse relationship between the contact angle and α. At higher α, the surface is extremely hydrophilic in character with a contact angle near zero. As α diminishes, hydrophobicity becomes dominant with the contact angle becoming greater than 140°. The halfway point of this trend is the wettability condition at α ≈ 0.219, which is the boundary between hydrophilic and hydrophobic regimes. These results highlight that relatively minor changes in interfacial energy can lead to major changes in wettability behavior. Therefore, precise control of α is of extreme importance in surface design and optimization for boiling heat transfer conditions.

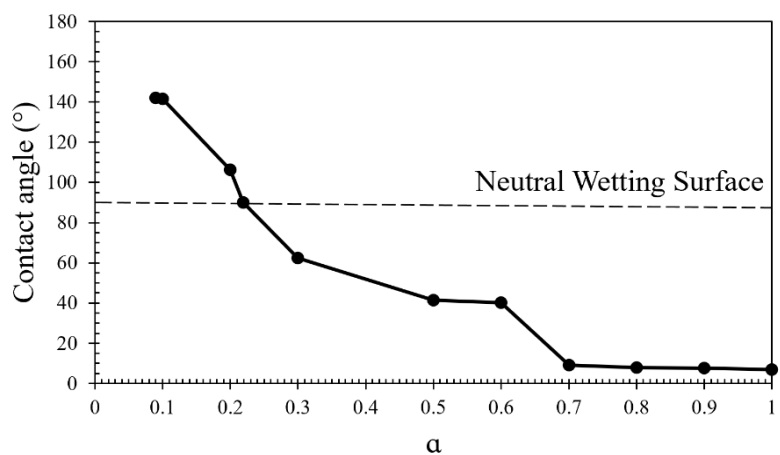

**Figure 6.** Measurement of water droplet contact angles on aluminum surfaces with a wide range of wettability.

**4-2 Boiling Behavior on Surfaces with Different Wettability**

In this section, evaporation and phase change behavior of boiling in thin water films on horizontal aluminum substrate is examined for interfacial energy ratios α = 0.700, 0.219, and 0.090. For α = 0.700, Figure 7 presents a series of snapshots that illustrate the transient phase change behavior of the liquid film at several boiling times.

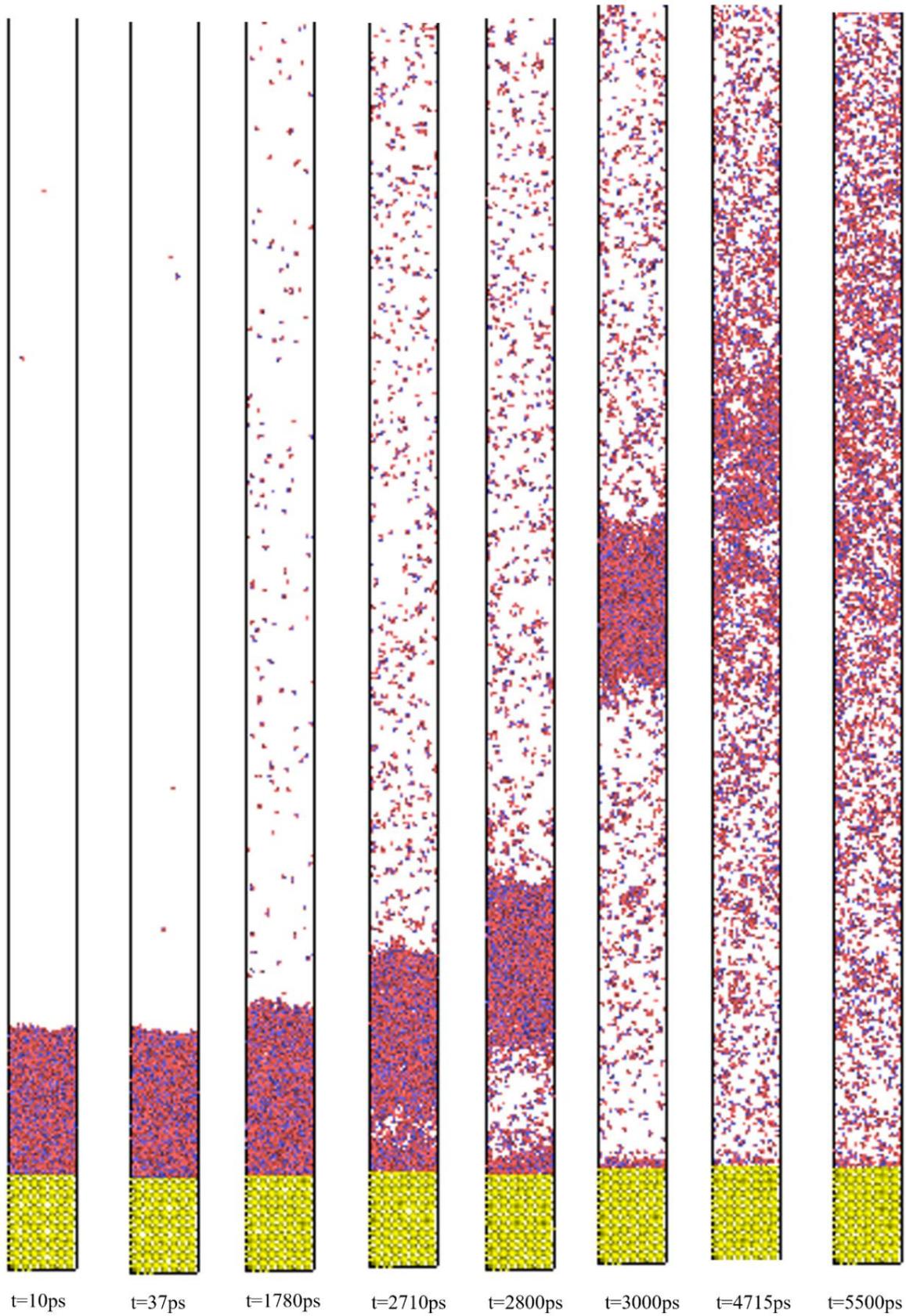

**Figure 7.** Snapshots of the phase transition process of a water film on an aluminum surface with a wettability coefficient of α = 0.7 at different boiling times.

The results indicate that as a rise in the temperature of the solid substrate takes place, the phase transition process moves from simple evaporation to explosive boiling. At initial periods of heating, surface evaporation dominates. As the temperature continues to rise, small bubble nuclei begin to develop within the liquid film, representing the initiation of nucleate boiling. With further augmentation of heating, bubbles grow and coalesce to create a persistent vapor layer. This film of vapor keeps on increasing by thermal expansion, forcing the remaining liquid clusters to move up, which characterizes the beginning of film boiling.

To better achieve the impact of surface potential energy on wettability and consequently on heat transfer through thin-film boiling, three surfaces with different ratios of solid-liquid interaction energy ($\alpha$ = 0.7, 0.219, and 0.09) were utilized, and their heat transfer characteristic was examined. To quantify the system's kinematic response, the displacement of the center-of-mass of the water film perpendicular to the aluminum substrate and its acceleration were calculated, as shown in Figure 8.

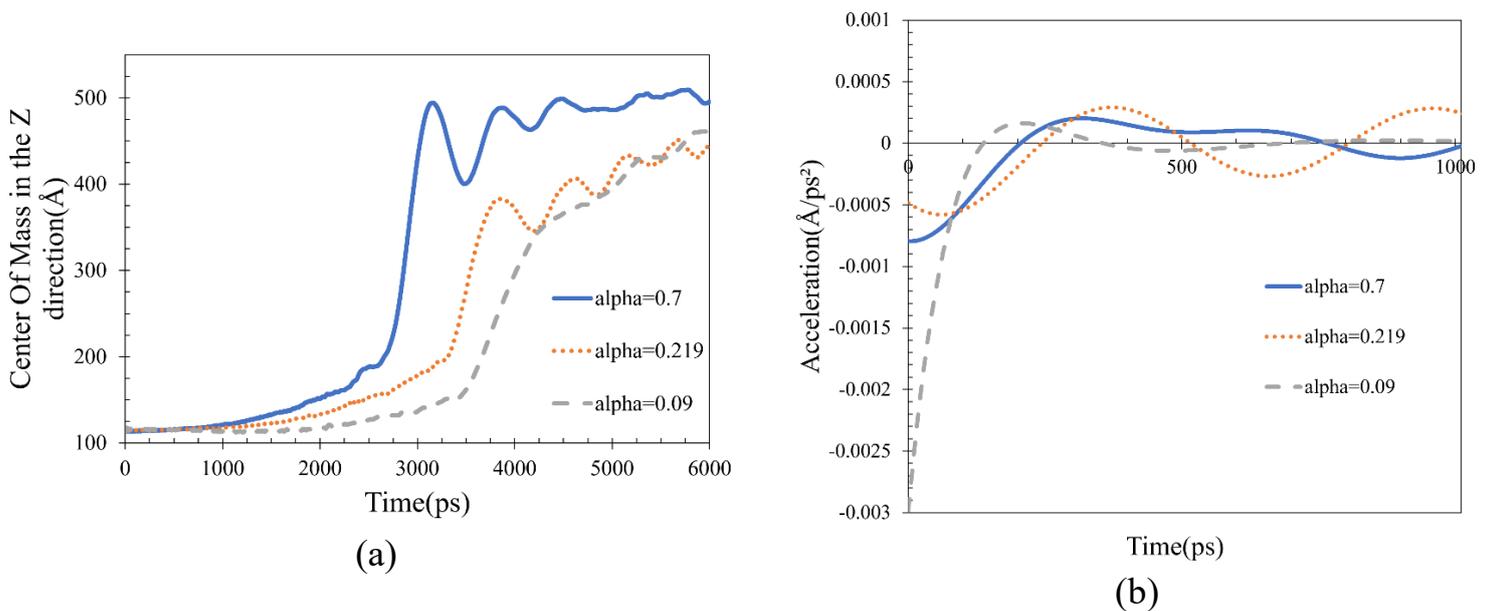

**Figure 8.** Analysis of water film dynamics: temporal variations of (a) the center-of-mass displacement and (b) atomic acceleration during the boiling process.

As figure 8a shows, upon the onset of heating, the center-of-mass of the water molecules in the vertical direction increases slowly at first primarily as a result of thermal expansion. Next, there is a point of inflection in the graph beyond which the height increases extremely sharply. This stage is simultaneous with the coalescence of the vapor cluster and the beginning of local vapor layer formation, which enhances the volumetric displacement of the liquid film in the upward direction. There follows oscillatory fluctuations due to the interaction of vapor pressure near and far from the surface.

As indicated by Figure 8b, to determine the onset time of interfacial instabilities, we followed a commonly cited criterion in literature that entails the selection of the first positive acceleration peak as a signal of the occurrence of nucleate boiling. The occurrence times of this maximum for the three surfaces were t≃200ps for the hydrophobic surface, t≃300ps for the hydrophilic surface, and t≃370ps for the neutral surface. Thus, though initial instabilities are achieved earlier on the hydrophobic surface (due to weaker adhesive forces and a lower energy barrier for forming vapor pockets), they fail to produce sustained volumetric displacement owing to poorer thermal coupling. Macroscopic jump in the center-of-mass follows later than that for the hydrophilic surface. In contrast, for the hydrophilic surface, heat transfer is enhanced and rewetting of the liquid following bubble bursting is stronger. Hence, following the maximum acceleration, the system enters the stable upward displacement regime sooner, ultimately reaching a higher terminal position.

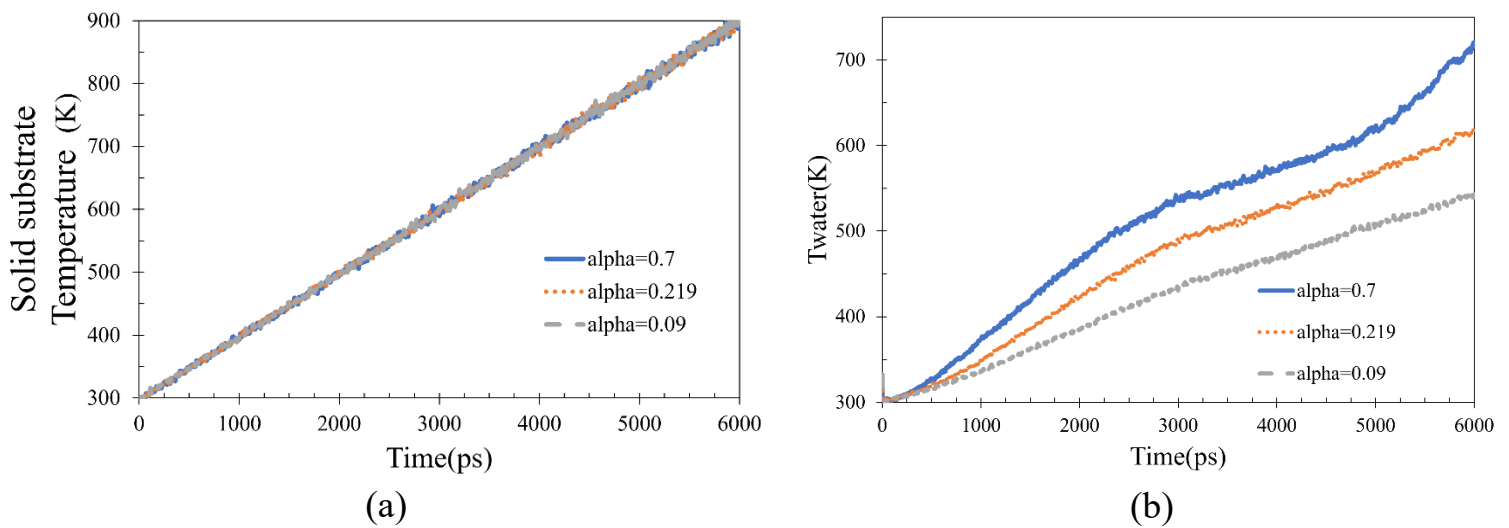

**Figure 9.** (a) Temperature evolution of the aluminum substrate during the boiling process. (b) Temperature evolution of the water film in contact with the substrate during the boiling process.

Figure 9 illustrates the substrate wall temperature and water film temperature for hydrophilic, neutral, and hydrophobic surfaces. The aluminum substrate was uniformly heated from 300 K to 900 K at a constant rate of 0.1 K·ps$^{-1}$ during the phase change process. As seen from Figure 9a, the temperature of the substrate showed almost linear and identical patterns under all three wettability states, demonstrating that the heat source in the solid layers behaved uniformly and surface wettability did not impact significantly on the temperature of the solid substrate itself.

As shown in Fig. 9(b), the temperature response of the water film is heavily influenced by the wettability of the surface. In this case, water molecules absorb energy from the solid substrate and gradually warm up. The temperature rise of the liquid results primarily from the increase in kinetic energy of the molecules as a consequence of solid surface impacts and interactions. Quantitative measurements of the heating rate of the liquid provide values of 0.064 K·ps$^{-1}$ for the hydrophilic surface, 0.048 K·ps$^{-1}$ for the neutral surface, and as low as 0.035 K·ps$^{-1}$ for the hydrophobic surface.

These findings indicate that on the hydrophilic surface, with a larger contact area and stronger interaction between water and solid, more energy is transferred to the liquid and thus the onset of nucleation of bubbles (ONB) at a lower temperature is easier. Conversely, on the hydrophobic surface, limited contact and enhanced interfacial resistance hinder the transfer of heat, resulting in a lower energy input by the water molecules and thus in the requirement for higher temperatures and longer times to initiate nucleation. The neutral surface shows an intermediate behavior.

Figure 10 also indicates the evolution of the number of evaporating water molecules of the three surfaces with different wettability. During the initial stage, the growth rate of evaporated molecules is moderate, which is indicative of surface evaporation. From a smooth rise, the nucleate boiling begins and accelerates as carbon dioxide emerges. As heat prolongs with increasing liquid temperature near the surface, the evaporation rate gradually grows. At about 2700ps for the hydrophilic surface, 3300ps for the neutral surface, and 3600ps for the hydrophobic surface, the slope of the curve suddenly takes on a sharp increase.

Such a sharp transition is an indication of vapor cluster coalescence, blanket formation, and migration towards film boiling. The hydrophilic surface with more efficient heat transfer and higher heating rates for the liquid has a vapor coalescence-threshold earlier, whereas the hydrophobic surface, having less efficient contact and lower energy flux to the liquid, experiences this transition later. When the increase is abrupt, most of the interfacial liquid is evaporated, and in the form of stabilized layer of vapor, the number of evaporated molecules increases at a diminishing rate.

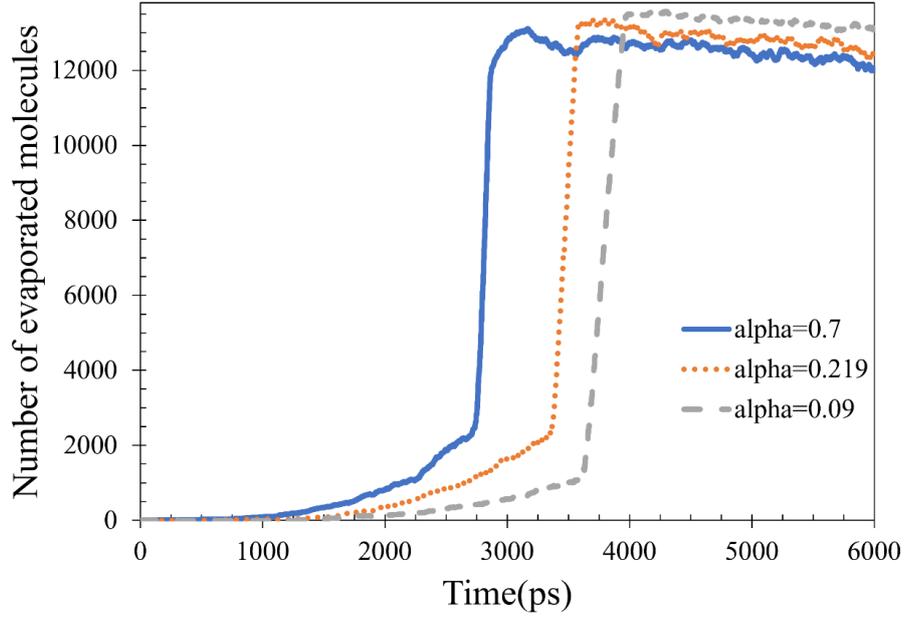

**Figure 10.** Variation of the evaporation ratio as a function of time for hydrophilic, neutral, and hydrophobic aluminum surfaces.

Heat flux variations can reveal the influence of surface wettability on the heat transfer characteristics during the boiling process. The heat flux is calculated using the following relation [1, 6, 7, 12, 16, 21, 24, 28-30, 36, 37, 39, 41-48]:

$$q = \frac{1}{A} \cdot \frac{\partial E_{water}}{\partial t} \tag{4}$$

Here, $E_{water}$ and A represent, respectively, the total internal energy of all water molecules and the surface area of the solid substrate. In this study, the effective heat transfer area was kept constant at A=16 nm² for all cases. The boiling curves corresponding to the three surfaces with different wettability coefficients are presented in Figure 11.

The overall trend in Fig. 11(a) follows the trend as described in the following descriptions. During the initial period (until ~100 ps), free convection leads to the rise of heat flux almost linearly. There are small differences between the three surfaces at this initial stage, and there is a slightly higher instantaneous flux for the hydrophobic surface due to interfacial temperature jumps.

With increasing time, the system starts the regime of enhanced nucleate boiling where the heat flux rises steeply until it reaches its peak. The values of critical heat flux (CHF) and the corresponding times are: $CHF_{\alpha=0.7} = 5216 \frac{MW}{m^2}$ at 1780 ps, $CHF_{\alpha=0.219} = 3979 \frac{MW}{m^2}$ at 2082 ps, and $CHF_{\alpha=0.09} = 2525 \frac{MW}{m^2}$ at 2618 ps.

Below the CHF point, the flux drops off, typical of the transition boiling as patches of vapor become more and more comprehensive in blanketing the surface.

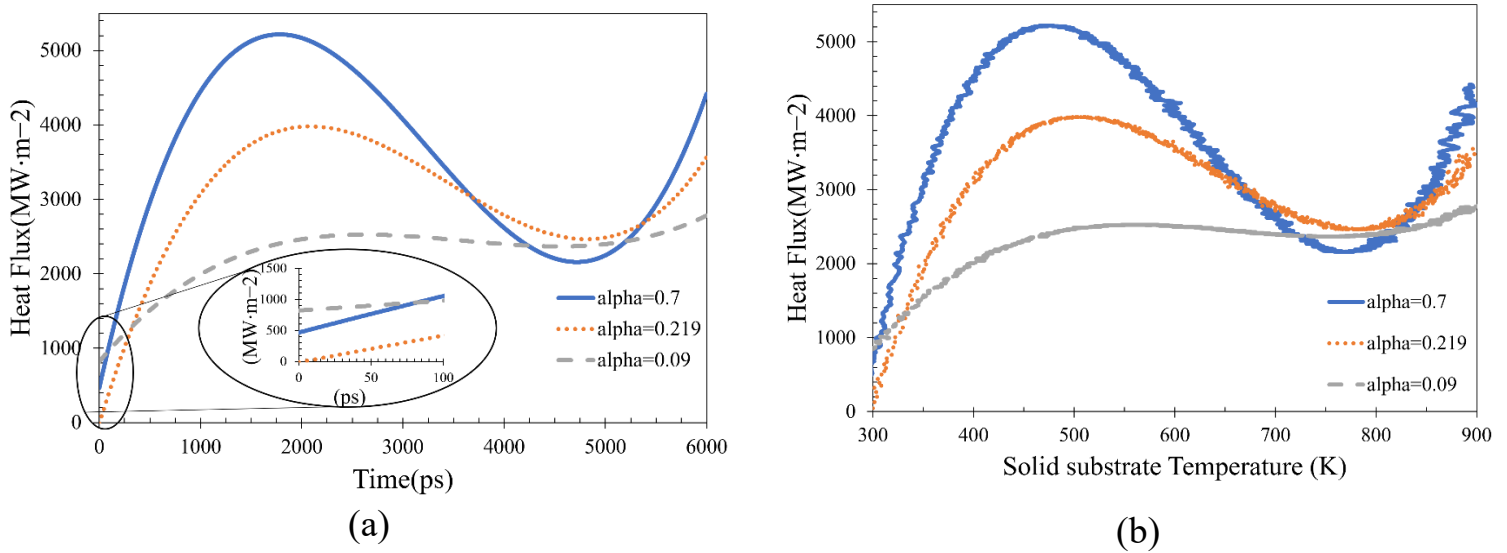

(a)                            (b)

**Figure 11.** Boiling curves for water films on aluminum surfaces with different wettability coefficients. (a) Temporal evolution of heat flux showing the transition from free convection to nucleate boiling, reaching the critical heat flux (CHF), and subsequently declining toward the minimum heat flux (MHF). (b) Corresponding variation of heat flux with wall temperature, highlighting the wall superheat associated with CHF and the Leidenfrost point.

With additional heating, the film boiling regime exists and the flux approaches its minimum or Leidenfrost point (MHF). In this stage, the liquid film separates from the hot wall and an equilibriating film of vapor is formed between the overheated surface and the liquid, through which heat is transmitted chiefly by the gaseous layer. Minimum heat flux (MHF) values and times are: $MHF_{\alpha=0.7} = 2157 \frac{MW}{m^2}$ at 4715 ps, $MHF_{\alpha=0.219} = 2463 \frac{MW}{m^2}$ at 4812 ps, and $MHF_{\alpha=0.09} = 2366 \frac{MW}{m^2}$ at 4533 ps. After crossing the Leidenfrost point, the flux grows moderately with higher wall superheating, but the increase is smaller than in the nucleate regime due to the thermal resistance of the vapor layer.

As can be seen from Fig. 11(b), the wall explosive boiling temperature at CHF (here treated as the explosive boiling temperature) is 466 K, 502 K, and 561 K for hydrophilic, neutral, and hydrophobic surfaces, respectively. The shift of the explosive boiling temperature is connected with solid-liquid interaction variations and intensity of kinetic energy exchange between the phases. Especially for the hydrophilic surface, increased solid-liquid coupling and improved water heating increase the kinetic energy acquisition of the water molecules, leading to a sooner critical regime at a higher CHF. On the hydrophobic surface, however, worse contact and less energy transfer demand increased substrate temperatures to attain similar boiling intensities, yet the CHF remains lower due to premature vapor patching and reduced liquid-solid contact. The intermediate behavior is expressed by neutral wettability. Wall temperatures at the MHF/Leidenfrost points are 767 K, 784 K, and 746 K for α = 0.7, 0.219, and 0.09, respectively. The trend reveals that the hydrophobic surface transitions to film boiling at the lowest wall temperature, the neutral surface possesses the largest value for the stabilization of vapor layer, and the hydrophilic surface is intermediate in between. Physically, high hydrophilicity enhances solid-liquid energy transfer, enabling higher CHF at lower explosive boiling temperatures; hydrophobicity weakens effective contact and initiates vapor layer inception at earlier stages, hence lowering the Leidenfrost temperature, while the neutral surface is a compromise between adhesion and volatility effects.

Heat transfer coefficient (HTC) is a highly significant parameter that characterizes the heat transfer rate. Enhancing HTC implies the ability to transfer more heat from the solid surface to the adjacent liquid film. In the present study, the interfacial heat transfer coefficient between the solid and liquid phases is calculated from the equation [6, 24, 28, 38, 42, 43, 46, 47]:

$$h = \frac{q}{T_{Al} - T_{water}} \qquad (5)$$

where $T_{Al}$ is the temperature of the aluminum substrate and $T_{water}$ is the temperature of the water molecules. The interfacial heat transfer coefficients of water molecules on aluminum surfaces with hydrophilic, neutral, and hydrophobic wettabilities are presented in Figure 12.

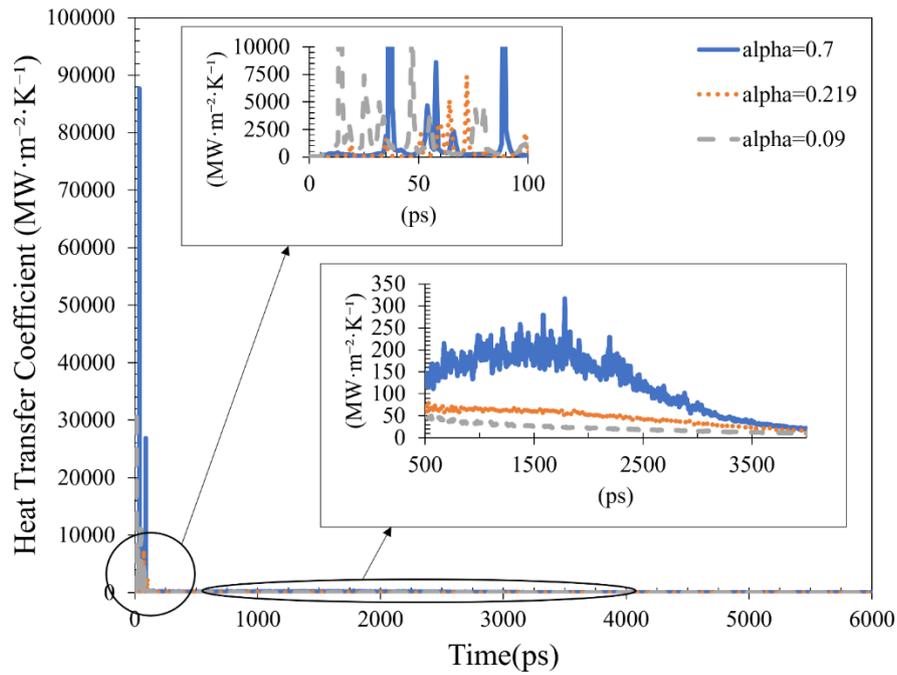

**Figure 12.** Time evolution of the heat transfer coefficient (HTC), with enlarged views highlighting the nucleate boiling and transition boiling regimes.

In the very early stage of the simulation, pronounced spikes in the heat transfer coefficient are observed. These transient surges originate from localized nucleation events occurring near the solid-liquid interface. On the one hand, the temperature difference between the solid and the liquid is still small, which amplifies the $\frac{q}{\Delta T}$ ratio; on the other hand, the onset of incipient nucleation at the interface leads to a momentary rise in heat flux. The combination of these factors results in abnormally high HTC values during the first few tens of picoseconds.

Subsequently, the HTC behavior clearly reflects the influence of surface wettability. For the hydrophilic surface, the HTC rapidly increases and fluctuates around a relatively high level (approximately 150-300 MW·m$^{-2}$·K$^{-1}$). The abundance of nucleation sites, faster rewetting after bubble departure, and efficient microlayer evaporation collectively sustain a high heat flux while suppressing the wall superheat, thereby maintaining a high HTC. In contrast, the neutral surface exhibits intermediate values (50-100 MW·m$^{-2}$·K$^{-1}$) with smoother fluctuations, indicative of reduced effective contact and lower nucleation frequency compared to the hydrophilic case. On the hydrophobic surface, the HTC remains at the lowest level (20-50 MW·m$^{-2}$·K$^{-1}$) and shows relatively uniform behavior, due to limited liquid-solid contact and enhanced contribution of heat conduction through a thin vapor film. At longer times (beyond ~2500 ps), the HTC for all three surfaces shows a decreasing trend.

This reduction is associated with progressive vapor coverage, bubble coalescence, and a shrinking effective liquid-solid contact area which is a phenomena that lowers the heat flux and enlarge the temperature difference across the interface. The decrease is earlier and more pronounced on the hydrophobic surface (due to its stronger tendency to form a stable vapor layer), whereas the hydrophilic surface, thanks to more efficient rewetting, sustains a higher HTC for a longer period. Eventually, as the system approaches the film boiling regime, the HTC of all surfaces converges to low values, reflecting the dominance of the thermal resistance of the vapor layer.

Extensive molecular dynamics (MD) simulations [1, 6, 7, 16, 22, 24, 28, 30, 37, 42, 45, 46, 48] have demonstrated that, at the solid-liquid interface, a thermal resistance exists that impedes the transfer of heat across the boundary. To quantify this effect, the interfacial resistance R against heat flux which is commonly referred to as the Kapitza thermal resistance, is defined as follows:

$$R = \frac{T_{AL} - T_{water}}{q} \qquad (6)$$

where $T_{AL}$ denotes the average temperature of the aluminum atoms and $T_{water}$ represents the average temperature of the water molecules adjacent to the surface. The overall variations of the Kapitza thermal resistance for the three different wettability conditions are illustrated in Figure 13.

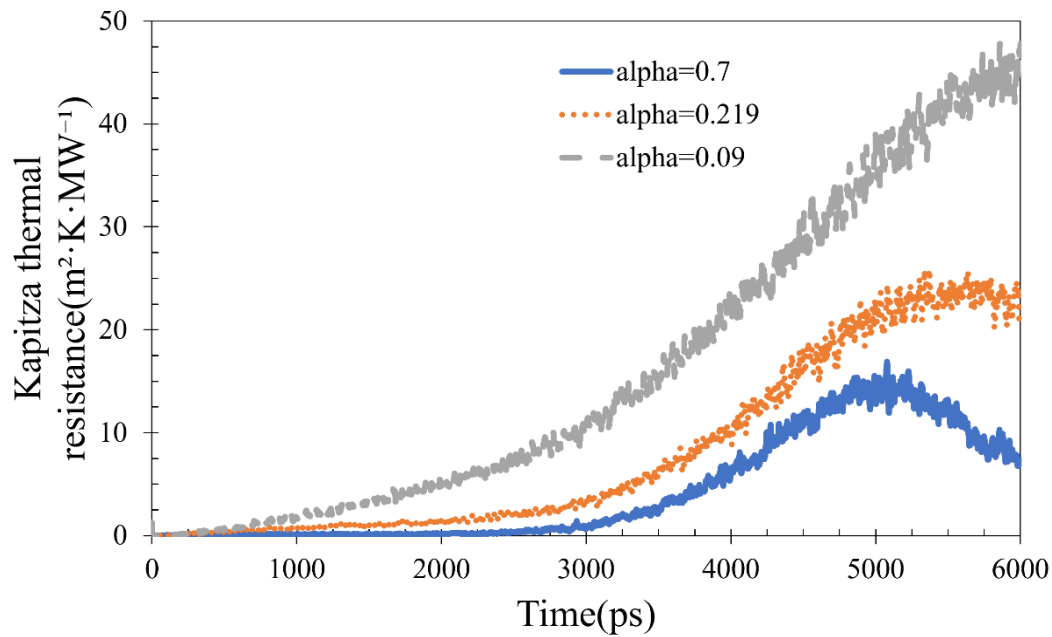

**Figure 13.** Time evolution of the Kapitza thermal resistance for three different wettability conditions.

Before the critical heat flux (CHF), interfacial thermal resistance on the three surfaces is small and increases only slowly, indicating that good liquid-solid contact is being maintained and kinetic energy exchange at the interface is efficient (active nucleate boiling, high heat flux, and small temperature difference). As CHF is exceeded and the system moves into the transition boiling regime, the vapor coverage of the surface improves, while the wall superheat and thermal resistance rise, and the heat flux falls. Hence, the thermal resistance rises more steeply.

In the stable film boiling regime (near the Leidenfrost point), the formation of the continuous insulating vapor layer dominates, and the thermal resistance reaches its maximum value. The ranking of resistance values between wettability conditions is as follows. Throughout the entire simulation, the hydrophilic surface has the lowest resistance and, even after the peak, experiences a modest decline due to localized rewetting. The neutral surface exhibits intermediate behavior, and the hydrophobic surface experiences the highest resistance as a result of less effective contact and more rapid vapor coalescence.

These results are consistent with the trends for the heat flux, interfacial heat transfer coefficient, and critical wall temperatures. Strong hydrophilicity leads to a higher CHF and delays the transition to film boiling by reducing interfacial resistance and enhancing heat transfer, whereas hydrophobicity strengthens thermal resistance, limits the efficiency of the energy exchange, and accelerates the onset of the film boiling regime.

## 5 Conclusions

For this study, non-equilibrium molecular dynamics (NEMD) calculations were performed to investigate pool boiling of ultra-thin water films on polished aluminum substrates with a particular focus placed on the surface wettability. An initial correspondence between the interfacial energy parameter ($\alpha$) and the equilibrium contact angle was established; lowering $\alpha$ from 1.0 to 0.09 converted the wettability from extremely hydrophilic to extremely hydrophobic, while neutral wettability existed at $\alpha \approx 0.219$. Cases such as hydrophilic, neutral, and hydrophobic, were afterwards quantitatively compared.

Kinematically, the first-peak time in film acceleration was around 200 ps, 300 ps, and 370 ps for hydrophobic, hydrophilic, and neutral surfaces, respectively. However, the macroscopic upward displacement of the film and the previously mentioned stable lift-off were observed for the hydrophilic surface, which is attributable to improved thermal coupling and faster rewetting upon bubble detachment. The average heating rates of the surface waters also confirmed the same pattern: 0.064, 0.048, and 0.035 K·ps$^{-1}$ for hydrophilic, neutral, and hydrophobic systems, respectively.

Boiling curves showed critical heat flux (CHF) of 5216, 3979, and 2525 MW·m$^{-2}$ for hydrophilic, neutral, and hydrophobic surfaces at 1780, 2082, and 2618 ps, respectively. Beyond the region of CHF, increasing vapor coverage and transition boiling produced decreasing heat flux, which decreased to minimum heat flux (MHF) at values of 2157, 2463, and 2366 MW·m$^{-2}$ at 4715, 4812, and 4533 ps, respectively. The corresponding wall temperatures at CHF were 466, 502, and 561 K, while the wall temperatures at MHF were 767, 784, and 746 K, respectively. Such thermal thresholds have a direct correspondence to the extent of kinetic energy exchange across the solid-liquid interface. Enhanced hydrophilicity enhances interfacial coupling, reduces the explosive boiling temperature, and raises the CHF, while hydrophobicity raises interfacial resistance, reduces effective contact, and facilitates film boiling transition.

The heat transfer coefficient (HTC) at the interface also exhibited consistent trends. After transient spikes during early nucleation, HTC reached the maximum for the hydrophillic, the intermediate value for the neutral, and the minimum value for the hydrophobic surface. Similarly, analysis of the Kapitza resistance showed the values to be low prior to CHF but rose quickly with the onset of transition boiling, reaching a peak in the film boiling regime. The hydrophilic surface consistently exhibited the lowest resistance, while for the hydrophobic situation, the most extreme values were found induced by faster coalescence of the vapor and reduced liquid-solid contact.

In principle, the findings indicate that wettability engineering aluminum surfaces to become stable hydrophilicity can decrease the explosive boiling temperature, significantly enhance the critical heat flux, and improve nanoscale thermal management performance.